\documentclass[referee]{aa}

\begin{document}
\thesaurus{02.16.2; 08.13.2; 08.13.1; 02.19.2}

\title {Sco X-1 and Cyg X-1: Determination
of Strength and Structure of Magnetic Field in the Nearest
Environment of Accreting Compact Stars}

\author{ Yu. N. Gnedin\inst{1} 
\and 
N. A. Silant'ev\inst{2}
\and
L. G. Titarchuk\inst{3}}

\offprints{Yu. N. Gnedin}

\institute{Main Astronomical Observatory of the Russian Academy of Sciences,
Pulkovo, St.Petersburg, 196140 Russia\\
e-mail: gnedin@gao.spb.ru
\and
Instituto Nacional de Astrof\'{\i}sica, Optica y Electronica,
Apartado Postal 51 y 216, C.P. 7200, Puebla, Pue., M\'exico\\
\and
George Mason University / CEOSR, Fairfax VA\\}

\date{Received....00, 2002; accepted....00,...}

\authorrunning{Yu. Gnedin, N. Silant'ev \& L. Titarchuk}
\titlerunning{Determination of magnetic field}
\maketitle

\begin{abstract}

We estimated the magnetic field strength of compact stars in X-ray
binaries Sco X-1 and Cyg X-1, via various methods of determination
of magnetic fields. For Sco X-1 we used three independent methods.
One of them is based on the correct account of the Faraday rotation of
polarization plane in the process of electron scattering of X-rays
from accreting neutron stars. Numerical calculations are made with
use of first X-rays polarimetric data presented by Long et al.\,
(1979). Other original method of determing the magnetic field
developed by Titarchuk at al. (2001) is based on observed 
quasi-periodic oscillations (QPO) frequencies in X-ray binaries 
that can be considered as magnetoacoustic oscillations of boundary 
layer near a neutron star. The optical polarimetric data obtained 
in 70-th have been also used for estimation of magnetic field of 
the neutron star in Sco X-1 and of nearest environment around 
the black hole in Cyg X-1.


\end{abstract}

\section{Introduction}

For X-ray sources of type Sco X-1 the Compton
scattering by electrons is very important for generation of the
radiation spectrum. The Compton scattering produces the
linear polarization of radiation from such type objects. The
polarization of X-ray sources have been first
considered by Rees (1975) and independently by Lightman \&
Shapiro (1975). They calculated the spectrum of polarized radiation
from an accretion disk.

But the spectrum of polarized radiation is drastically changed due to
the Faraday rotation of the polarization plane in a magnetic
field. Gnedin \& Silant'ev (1980, 1997) (see also the book by
Dolginov et al. 1995) have calculated the spectrum of polarized
radiation for the various astrophysical objects with magnetic
fields taking into account the Faraday rotation effect and
suggested the new method of determination of stellar magnetic
field with the spectrum of broad band linearly polarized radiation
(see, for example, Gnedin \& Silant'ev 1984; Gnedin et al.
1988). The application of their method to accretion disks around
supermassive black holes (AGNs, QSOs) has been recently developed
by Gnedin \& Silant'ev (2002).

The first rough estimation of a magnetic field of Sco X-1, based on
the first X-ray polarimetric observation made by Long et al. (1979),
was done by Gnedin \& Silant'ev (1980).

Recently Titarchuk et al. (2001) presented the detail model of 
quasi-periodic oscillations (QPO)
of Sco X-1 using a large set of Rossi X-ray Timing Explorer (RXTE)
data for this object. The QPOs have been observed not only into
kHz range, but also at ~6 Hz. There is an almost linear
correlation between Sco X-1's kHz and 6 Hz QPOs. They identify the
~6 Hz frequencies as acoustic oscillations of a Keplerian disk
around the neutron star (NS) that is formed after radiation
pressure near the Eddington accretion rate destroys the disk. They
showed that the disk is placed near one NS radius from the surface
of NS. It is the remarkable fact that they could estimate the Sco
X-1 magnetic field. It appears to be $ 0.7\times10^{6} G $ at
about one NS radius above the NS surface.

Let us remind also that still Gnedin \& Shulov (1971) and Nikulin
et al. (1971) claimed some evidence of circularly polarized light
of Sco X-1 in optical spectral range. Though the followed
measurement (see Illing \& Martin 1972) did not confirm the
existence of circular polarization for Sco X-1, Kemp et al. (1972)
claimed the possible existence of the strong nonstationary circular
polarization state. They called this state as a "flaring
polarization". Nevertheless the situation with determination of
magnetic field of Sco X-1 was unclear till now. The last
results by Titarchuk et al. (2001), allow  to reconsider this
situation and to compare the results of various methods of
magnetic field estimations. We start with the detail description
of presented methods of magnetic field estimations of
compact stars emitting X-rays.

\section{Linear Polarization Spectrum of X-ray Radiation from Magnetized
 Disks and Envelopes Around a Compact Star}

The X-ray radiation of a compact star (neutron star, black hole),
scattered on electrons of the envelope of an accreting matter,
acquires linear polarization. It is well known, however, that
integral polarization from a spherical envelope is equal to zero
due to total compensation of directions of electric vectors from
various areas of this envelope. If magnetic field is absent, the
nonzero net polarization requires nonspherical shape of the scattering
electron envelope. If there is the presence of a magnetic field in
the circumstellar electron envelope, the situation is radically
changed. The scattered radiation undergoes the Faraday rotation of its
polarization plane due to the propagation in a magnetized plasma.

The angle of the Faraday rotation $ \chi $ is determined by the
expression (see Gnedin \& Silant'ev 1980, 1997)

\[
\chi = \frac{1}{2} \delta \tau_T \cos{\theta},
\]
\begin{equation}
\delta =
\frac{3 \omega _{B} c}{2r_{e}\omega^2} \cong 1.2
\left(\frac{B}{10^{6}G}\right)\left(\frac{1 keV}{\hbar
\omega}\right)^{2},
\label{1}
\end{equation}

\noindent where $ \tau _{T} $ is the Thomson optical thickness of the envelope, $
\theta $ is the angle between the directions of a magnetic field ${\bf B}$ and
radiation propagation ${\bf n}$, $ r_{e} = e^2 / m_{e}c^2 $ is 
classic electron radius, $ \omega _{B} = eB / m_{e}c $ is the
cyclotron frequency.

If the magnetic field increases, the angle $ \chi $ increases
too and partly polarized scattered radiation begins to undergo the
Faraday rotation. The rotation angles $ \chi $ are different for
waves scattered in different volumes along the line of sight 
${\bf n}$, and, as a result, the total radiation from all volume
elements will be depolarized (see Fig.\,10 from the review by Gnedin
\& Silant'ev 1997).

The spectrum of polarized radiation from a spherically symmetric
envelope has been calculated by Gnedin \& Silant'ev (1980). For a
spherical envelope with the dipole magnetic field only the
magnetic equator volumes don't produce the Faraday rotation of
polarization plane because of for these volumes $
\cos{\theta}\approx 0 $ and $\chi \approx 0$. As a result, only the 
radiation scattered in equatorial volumes acquires net polarization 
in the plane (${\bf nM} $), where ${\bf n}$ and ${\bf M}$ 
are the directions of line of sight and dipole magnetic field,
correspondingly. Namely this radiation alone gives rise to
non-zero integral polarization from the magnetized envelope. The
shape of the spectrum of the integral linear polarization
has a bell-shaped form. Indeed, for very short wavelengths the
Faraday rotation angles tend to zero and the integral polarization
disappears due to spherical symmetry of the envelope. For very long
wavelengths the angles of Faraday's rotation increase and the 
near equatorial region with $\chi \approx 0$ decreases what 
diminishes the integral polarization. So, the maximum integral
polarization corresponds to such magnetic field strengths and
wavelengths when the mean rotation angle $ \chi \approx 1 $. The
process of strong depolarization begins with the the values of
magnetic field and wavelengths when $ \delta \tau _T > 1 $ (see
Eq.\,(1)).

For disk-like electron envelope without the true absorption 
with $\tau _T \gg 1$ we have for outgoing radiation (see Silant'ev 2002):

\begin{equation}
P_{l}({\bf n},{\bf B})\cong 0.914\,\frac{1-\mu^2}{1+2\mu}\,\frac{1}{\sqrt{1+\delta^2
\cos^{2}\theta}},
\label{2}
\end{equation}

\noindent where $ \mu = \cos{\vartheta} $, and $ \vartheta $ is the angle
between the line of sight $\bf n$ and the normal to the disk surface.
The terms without the square root describe roughly the polarization degree
$P_l({\bf n})$ of radiation in the absence a magnetic field. So, the formula 
can be written as

\begin{equation}
P_{l}({\bf n},{\bf B})\cong \frac {P_{l}({\bf n})}{\sqrt{1+\delta^2\cos^{2}
\theta }}. 
\label{3}
\end{equation}

\noindent Later we shall use this formula for the estimation of magnetic field 
in the environment of the source Sco X-1.

For the spherically symmetric optically thin envelope around the
neutron star with the dipole magnetic field the degree of linear
polarization of X-rays scattered into this envelope can be
estimated by the expression

\begin{equation}
P_{l}({\bf n},{\bf B})=\tau _{T}f(R_{0}/R_{s};\, \delta _{s}\tau _{0};\,
\theta)\sin^{2}{\theta }.
\label{4}
\end{equation}

\noindent Here $ R_{0}/R_{s} $ is the ratio of the envelope radius to the
neutron star radius, $ \delta _{s} $ is the depolarization
parameter (1) that corresponds to the stellar equatorial magnetic
field strength. The function $ f $ is tabulated by Gnedin \&
Silant'ev (1980, 1984).

In the region of the depolarization when $ \delta _{s} \tau _{T} >
1 $ the asymptotic expression of the function $ f $ takes a place:

\begin{equation}
f\approx (\delta _{s}\tau _{T})^{-\frac{n-1}{n+2}}\sim(\omega
/\omega_{0})^{\frac{2(n-1)}{(n+2)}}.
\label{5}
\end{equation}

\noindent Here $n$ is the exponent of radial dependence of the electron density
in the envelope: $ N_{e} \sim r^{-n}\quad (n\neq 0) $. The value $
\omega _{0}=(3\omega _{B}c/2r_{e})^{1/2} $. If, for example, $ n=2
$, the function $ f $ drops with frequency decrease as $ f\sim
\omega^{1/2} $. For $\delta_{s} \tau_{T}\ll 1$ (large frequencies)
the spectrum of polarization degree has universal dependence: 
$P_{l}\sim \omega^{-4}$.

\section{Magnetic Field Estimation from Magnetoacoustic 
Oscillations in Neutron Star Binaries}

The original method to determine the magnetic field around
neutron stars based on observed kHz and viscous 
quasi-periodic oscillations (QPO) frequencies
has recently been proposed by Titarchuk et al. (2001). They have
analyzed magnetoacoustic wave formation on the layer between a
neutron star surface and the inner edge of a Keplerian disk and
derived formulas for the magnetoacoustic wave frequencies for
different regimes of radial transition layer oscillations. As a
result they demonstrated that one can use the QPO as a new kind of
probe to determine the magnetic field strengths of neutron stars
in the binaries.

For the two extreme (acoustic and magnetic) cases Titarchuk et
al. (2001) have been obtained an approximate formula for the
magnetoacoustic frequency $\omega_{MA}$:

\begin{equation}
\omega_{MA}\approx
\left\{\frac{\left(\frac{\beta_{S}}{\pi}\right)^{2}V^{2}_{S}}
{4(r_{out}-r_{in})^{2}}+
\frac{\left(\frac{\beta_{M}}{\pi}\right)^{2}\left(\frac{\alpha
+2}{4}\right)^{2}V^{2}_{A}(r_{out})}{\left[r_{out}-r_{in}
\left(\frac{r_{in}}{r_{out}}\right)
^{\frac{\alpha}{2}}\right]^{2}}\right\}^{1/2}.
\label{6}
\end{equation}

\noindent Here $ V_{S} $ is the sound velocity, $ V_{A} $ is the Alfven
velocity, $ r_{out} $ and $ r_{in} $ are the outer and the inner
radii of a transit layer. The index $ \alpha $ is related to the
multipole magnetic field through the expression of $ V^2_A \sim
r^{-\alpha} $, so that $ \alpha =6, 8, 10 $ are for the dipole,
quadrupole, and octupole, respectively.

The parameters $ \beta_{S} $ and $ \beta_{M} $ are determined by
the transcendental equation (see Titarchuk et al. 2001):

\begin{equation}
\tan{\beta}=-\frac{2(\alpha-2)\beta}{(\alpha
+2)\left\{\frac{\eta\beta^2}{(\eta -1)^2}+\left[\frac{\alpha
-2}{\alpha +2}\right]^2\right\}},
\label{7}
\end{equation}

\noindent where $ \beta =z_{out}-z_{in} $ and $ \eta
=z_{out}/z_{in}=(r_{out}/r_{in})^{(\alpha +2)/2} $, $ z_{out} $
and $ z_{in} $ are z-axis for the transit layer.

Titarchuk et al. (2001) estimated approximately the values of $
\beta_{S} $ and $ \beta_{M} $:

\begin{equation}
\beta_{M}\approx\frac{\pi}{2}+\frac{2}{\pi}\left[\frac{(\pi/2)^2\eta}
{(\eta-1)^2}+\frac{1}{4}\right]
\label{8}
\end{equation}

\noindent for $ \alpha =6 $ and

\begin{equation}
\beta_S \approx \left\{1.5 /\,[1+1.5\eta/(\eta-1)^2]\right\}^{1/2}
\label{9}
\end{equation}

\noindent for $ \alpha=0 $ (the pure acoustic case).

Substituting the sound velocity $ V_S=(kT/m_p)^{1/2} $ and the
Alfven velocity $ V_M=B(r_{out})/(4\pi\rho)^{1/2} $ one can
estimate the magnetic field strength at the outer boundary of the
transition layer. Then using the multipole law for the magnetic
field one can also estimate the magnetic field strength at the
neutron star surface.

\section{Magnetic Field Estimation from Circular Polarization of 
Optical Light of a Binary System}

The broad-band polarimetry of the optical continuum radiation is the
effective direct method of the magnetic field estimation. The
electromagnetic wave which is incident onto plasma with a magnetic
field produces oscillations of electron velocity. As a result an
additional Lorentz force appears:

\begin{equation}
{\bf F}={\bf F}\left(\frac{\omega_B}{\omega},\,{\bf E}\times 
{\bf B}\right),
\label{10}
\end{equation}

\noindent which depends on the ratio of the cyclotron frequency 
to radiation one and on the angle between the directions of wave's 
electric vector ${\bf E}$ and the magnetic field ${\bf B}$. As a result
of the action of this Lorentz force magnetized plasma acquires
dichroism and birefringence properties, becoming similar to any
anisotropic medium. In a magnetized plasma two types of
electromagnetic waves (normal modes or normal waves) with
different types of elliptical polarization should be propagated.
These normal waves are usually called ordinary (O.W.) and
extraordinary (E.O.W.) with their intrinsic refraction indices,
phase velocities and polarization. O.W. behaves as an usual
electromagnetic wave in a plasma without a magnetic field. For
E.O.W. transport coefficients for various emission
(magnetobremsstrahlung) and scattering (electron scattering)
processes have resonance at cyclotron frequency $ \omega_B $ (in
detail see Dolginov et al. 1995).

In the case when the radiation frequency $ \omega $ is much larger
than the cyclotron frequency $ \omega_B $, the radiation is
predominantly circularly polarized:

\begin{equation}
P_V=2(\omega_B/\omega)\cos{\theta};\quad P_l\sim P_v^2\sim
(\omega_B/\omega)^2.
\label{11}
\end{equation}

One of the most important case is the cyclotron resonance: $
\omega \approx\omega_B $ (remember that $\omega_B/\omega \simeq 
0.93\cdot 10^{-8}\lambda(\mu m)B(G)$ ). For the optical range this case
corresponds to magnetic field strengths $ B\sim 10^7\div 10^8 G $.
At cyclotron resonance the plasma radiation is completely
polarized:

\begin{equation}
P_l=\frac{\sin^2{\theta}}{1+\cos^2{\theta}};\quad
P_V=\frac{2\cos{\theta}}{1+\cos^2{\theta}};\quad P_l^2+P_V^2=1.
\label{12}
\end{equation}

In our case of the close binary system with a magnetized neutron
star the optical radiation of the accreting disk or plasma
envelope around a neutron star can be circularly polarized. As a
result the estimation of the neutron star magnetic field strength
can be obtained from the measured optical circular polarization.

\section{Sco X-1: Estimation of Magnetic Field Strength}

The X-ray polarization of Sco X-1 was searched by Long et al.
(1979). Multiple observations of Sco X-1 were carried out with the
Bragg crystal polarimeter aboard OSO8. An unambiguous detection of
the time-averaged polarization was obtained for Sco X-1 which has
a polarization $ P_l=(0.39\pm 0.20)\% $ at 2.6 keV and $
P_l=(1.31\pm 0.40)\% $ at 5.2 keV. These data give a signature of
the polarization fall at energy values below 5.2 keV. If one
suggests that this fall is due to the Faraday depolarization
effect one can estimate the magnetic field strength at an environment
of the neutron star in Sco X-1 binary system. The ratio $ P_l(2.6
\, keV)/P_l(5.2\, keV)\approx 0.3 $.
 
Using  Eq.\,(3), we obtain for the 5.2 keV
case $\delta \cos{\theta} \simeq \sqrt{(P_l({\bf n})/1.31)^2-1} $. 
The choice of $P_l$ depends on of the angle of inclination of a disk. 
It is clear that we are to take $P_l\ge 1.31\%$. As a result we have the
estimation:
 
\begin{equation}
B \ge 2.3\cdot 10^{7}\sqrt{(P_l/1.31)^2-1} G.
\label{13}
\end{equation}

\noindent This estimation takes place for the model of optically thick 
inclined disk. If we choose $P_l\simeq 2$, we obtain $B\ge 2.6\cdot 10^7$\,G.
Analogous estimation for the 2.6 keV case gives $B\ge 2.8\cdot 10^7$\,G, i.e.
in range of observational errors both estimations coincide one with another. It
seems the polarized radiation forms at the inner part of accretion disk where
the temperature is higher.

Another estimation follows from the model of spherical electron envelope in
dipole magnetic field. Considering that the polarization 1.31\% corresponds to
the position at the polarization spectrum curve which is near maximum, i.e.
$\delta_s\tau /\eta^3 \sim 1$, we obtain

\begin{equation}
B_s\ge 2.3\cdot 10^7\, \frac{\eta^{3}}{\tau}\,G.
\label{14}
\end{equation}

\noindent Here, $\eta =R_0/R_s$ is the ratio of the envelope radius to 
the neutron star radius and $\tau $ is the Thomson thickness of the envelope. 
Taking $\eta \simeq 2$ and $\tau \simeq 0.25$, we obtain 
from (14) the estimatiom $B_s\simeq 7\cdot 10^8$\,G, or the value
$B\simeq 10^8$\,G in the region of an envelope. The estimations (14) 
considerably greater than those from (13). In reality this difference
is more considerable because the condition 
$P_{l}(5.2 keV)/P_{l}(2.6 keV)\simeq 3.3$
can be carried out only if $\delta_s > \eta^3 /\tau$.

Another way of estimation of Sco X-1 magnetic field is to use
observed kHz and viscous QPO frequencies. This method was
developed by Titarchuk et al. (2001). They used the best-fit
parameters of Sco X-1 transition layer and determined the magnetic
field strength $B_{TL}=(1.0\pm 0.05)\cdot 10^6 \,G$ and 
$R_{TL}=2.12\cdot 10^6$ cm in the transition layer. An
extrapolation of the magnetic field from $R_{TL}$ towards the
neutron star radius gives us $B_s=8\cdot 10^6 \,G$ and
$B_s=3.3\times 10^7 \,G$ for the dipole and octupole fields,
respectively. These values appear close to the X-ray
polarimetric estimation of neutron star magnetic field given by
(13).

Third way of estimation of Sco X-1 magnetic field is connected
with observations of circular polarization of optical light in Sco
X-1 binary system.

Gnedin and Shulov (1971) and Nikulin et al. (1971) claimed
occasional appearance of appreciable and variable circular
polarization of optical light of Sco X-1 (see, also, Severny \&
Kuvshinov, 1975). Though Illing \& Martin (1972) didn't confirm
these results, Kemp et al. (1972) claimed also evidence of the
occasional appearance of circular polarization of Sco X-1 that was
named by them as "flaring polarization".

The net circular polarization of Sco X-1 was estimated with
Eq.\,(11). Gnedin \& Shulov (1971) made the observations of
circular polarization in the yellow-red light that corresponds the
effective wavelength $\lambda_{eff}=0.64 \mu m$. For their case
estimation (11) give the following value of the magnetic field
strength:

\begin{equation}
B\geq 2\cdot 10^6 \left(\frac{P_V}{0.01}\right)\,G.
\label{15}
\end{equation}

\noindent This value corresponds quite well to the magnetic field magnitude
derived by Titarchuk et al. (2001). However, the magnitude (11)
should be considered only as the low limit for the real magnetic
field of the neutron star because one need to take into account
the contribution of the optical companion in the radiation of the
accreting disk of Sco X-1 system. Therefore the real value of the
neutron star magnetic field must be increased at least an order
and reaches the value at the level $\geq 10^7\,G$.

This is the remarkable thing that three independent estimations
give close values for the magnetic field
strength in Sco X-1 system.

It should be noted that the case of accretion disk with the chaotic (turbulent)
magnetic fields gives the same estimation (13) as for some regular magnetic
field with appropriate choice of the angle $\theta$ (see formulas in Silant'ev,
2002). Besides, the model with chaotic magnetic fields explains naturally the
"flaring" character of circular polarization that is due to the fluctuations
of a magnetic field.

\section{Cyg X-1: Estimation of Magnetic Field Strength in a Plasma 
Near the Black Hole}

How to extract energy from a rotating black hole is important
issue in astrophysics. Many people have considered various
alternative mechanisms for extracting energy from a rotating black
hole. Among them the most promising one is the famous
Blandford-Znajek mechanism (Blandford \& Znajek 1977). In this
mechanism a Kerr black hole is assumed to connect with
surrounding matter with magnetic field lines. The magnetic field
lines thread the black hole's horizon and its rotation twists the
magnetic field lines and transports energy and angular momentum
from the black holes to the accretion disk (Blandford 2001; Li
2002; Li \& Paczynski 2000, and references therein).

A magnetic field connecting a black hole to a disk has important
effects on the balance and transfer of energy and angular
momentum. The global structure of black hole magnetospheres
involving axisymmetric magnetic field and plasma injected from an
accretion disk has been extensively investigated for explaining
various observational features of transient X-ray binaries and
active galactic nuclei (see, e.g., for the review by Beskin 1997).

Recently Robertson \& Leiter (2001) claimed an evidence for
intrinsic magnetic moments in black hole candidates.

The Electro-Magnetic Black Hole theory has been recently developed
by Ruffini (2002). He has applied this theory to the analysis of
the GRB phenomenon. He showed the structure of burst and afterglow
of GRB can be explained within the theory based on the vacuum
polarization process occurring in an Electro-Magnetic Black Hole.

Unfortunately there is no till now the direct evidence of the
presence magnetic field in near environment around a black hole.

The marginal evidence of X-ray polarization of Cyg X-1 allows in
the accretion disk around the compact X-ray object (black hole)
(Weisskopf et al. 1977, Long et al. 1980). Multiple observations
of Cyg X-1 were carried out by Long et al. (1980) with the Bragg
crystal polarimeters aboard OSO8. The detection of the
time-averaged polarization was obtained for Cyg X-1 at the level:
$P_l(2.6\, keV)=2.4\%\pm1.1\%$ and $P_l(5.2\, keV)=5.3\%\pm2.5\%$.
If the decrease of polarization at 2.6 keV
one explains as a result of the Faraday depolarization, the estimation
of a magnetic field strength via Eq.\,(2) gives $B \geq 10^7 G$.
It is curious that this magnetic field magnitude is quite
sufficient to explain the famous result of measurement by
Michalsky \& Swedlund (1977) of variable optical circular
polarization of Cyg X-1: $P_V\cong 5\cdot 10^{-4}$.

\section{Conclusion}

We estimated the magnetic field magnitudes of two X-ray binaries:
Sco X-1 and Cyg X-1. We used for estimation two different methods.
One of them is connected with correct account of the Faraday rotation
of polarization plane in the process of scattering of X-ray
radiation (method developed by Gnedin and Silant'ev 1980,\,1984,\,1997).
Another original method of determining the magnetic field of the
objects like Sco X-1 is developed by Titarchuk et al. (2001). This
method is based on observed quasi-periodic oscillations (QPO) frequencies 
in X-ray binaries and their interpretation as magnetoacoustic oscillations 
in neutron star binaries. It is remarkable that both methods give close
results: magnetic field strength of neutron star in Sco X-1
appears at the level of $B_S\approx 10^7\,G$. It is wonderful
that measurements of optical circular polarization of Sco X-1 give
the same estimation of neutron star magnetic field in Sco X-1 (see
Eq.\,(15) with taking into account the fact that optical radiation
is generated in outer layers of the accretion disk comparatively
far away from the neutron star itself).

\begin{acknowledgements}

Gnedin is  grateful for support from Russian
Research Program: "Nonstationary Phenomena in Astronomy" and
program "Astronomy" of Russian Department of Science and
Technology.

\end{acknowledgements}

\end{document}